\documentclass[preprint,showpacs,amsmath,amssymb]{revtex4}

\usepackage{graphicx}
\usepackage{dcolumn}
\usepackage{bm}

\begin{document}

\title{Phenomenological study on the $\bar p N\to \bar NN\pi\pi$ reactions}

\author{Xu Cao$^{1,3,5}${\footnote{Electronic address: caoxu@impcas.ac.cn}} }
\author{Bing-Song Zou$^{2,3,4}$}
\author{Hu-Shan Xu$^{1,3,4}$}

\affiliation{$^1$Institute of Modern Physics, Chinese Academy of
Sciences, Lanzhou 730000, China\\
$^2$Institute of High Energy Physics, Chinese Academy of Sciences,
Beijing 100049, China\\
$^3$Theoretical Physics Center for Sciences Facilities, Chinese
Academy of Sciences,
Beijing 100049, China\\
$^4$Center of Theoretical Nuclear Physics, National
Laboratory of Heavy Ion Collisions, Lanzhou 730000, China\\
$^5$Graduate University of Chinese Academy of Sciences, Beijing
100049, China}

\begin{abstract}
We extend our recent phenomenological study of $pN \to NN\pi\pi$
reactions to the $\bar{p}N \to \bar{N}N\pi\pi$ reactions for
anti-proton beam momenta up to 3.0 GeV within an effective
Lagrangian approach. The contribution of $N^*(1440)$ with its
$N\sigma$ decay mode is found to be dominant at the energies close
to threshold for $\bar NN\pi^+\pi^-$ and $\bar NN\pi^0\pi^0$
channels. At higher energies or for $p\bar n\pi^-\pi^-$ and $\bar
NN\pi^\pm\pi^0$ channels where $N^*(1440)\to N\sigma$ mode cannot
contribute, large contributions from double-$\Delta$, $\Delta(1600)
\to N^*(1440)\pi$, $\Delta(1600) \to \Delta\pi$ and $\Delta(1620)
\to \Delta\pi$ are found. In the near-threshold region, sizeable
contributions from $\Delta \to \Delta\pi$, $\Delta \to N\pi$, $N \to
\Delta\pi$ and nucleon pole are also indicated. Although these
results are similar to those for $pN\to NN\pi\pi$ reactions, the
antinucleon-nucleon collisions are shown to be complementary to the
nucleon-nucleon collisions and may even have advantages in some
aspects. The PANDA/FAIR experiment is suggested to be an excellent
place for studying the properties of relevant $N^*$ and $\Delta^*$
resonances.
\end{abstract}
\maketitle{}

\section{INTRODUCTION}

As an interesting field to study baryon spectrum and properties of
strong interaction, double pion production in pion-, photo- and
electro-induced reactions has been extensively
explored~\cite{reviewlee}. Recently, we have performed a
comprehensive theoretical analysis of the double pion production in
nucleon-nucleon collisions~\cite{caoprc} based on the new data from
CELSIUS and COSY experiments in the past few years~\cite{ppdata}. It
is meaningful to extend the study to the closely related $\bar p
N\to \bar NN\pi\pi$ reactions, and herein we present the results.

The experimental studies on the $\bar p N\to \bar NN\pi\pi$
reactions were mainly performed in the years around
1970~\cite{ope,regge,roperargue,Mason,lysdata,eastdata,lbbook} with
some additions on the $\bar{p}p \to \bar{p}p\pi^+\pi^-$ channel by
the JETSET Collaboration at 1997~\cite{JETSET}. The data were still
scarce. On theoretical side, the one pion exchange (OPE)
model~\cite{ope} and Regge pole model~\cite{regge}, focusing on the
beam momenta above 3.0 GeV, included the double-$\Delta$ excitations
only. However, on experimental side, there was an argument about the
data of $\bar{p}p \to\bar{p}p\pi^+\pi^-$ at the beam momenta around
3.0 GeV~\cite{roperargue} whether there was contribution from a
$N^*$ with mass about 1400 MeV and width about 80 MeV, respectively.
Also, the experiment of $\bar{p}p \to\bar{p}p\pi^+\pi^-$ at the beam
momentum of 2.5 GeV~\cite{Mason} claimed an enhancement at a
$\Delta\pi$ invariant mass of 1370 MeV. From the modern point of
view, these data might show the presence of the Roper resonance
$N^*(1440)$ in the $\bar{p}p \to \bar{p}p\pi^+\pi^-$ channel. As a
matter of fact, the $N^*(1440)$ resonance should play essential role
in this channel, which can be postulated from our analysis of $NN\to
NN\pi\pi$ reactions~\cite{caoprc,Alvarez} where the $N^*(1440)$ was
found to be important in the near-threshold region. The
$\Delta(1600)$ and $\Delta(1620)$ resonances are also expected to
show up in the $\bar p N\to \bar NN\pi\pi$ reactions at high
energies because they are found to be important to describe the data
of $NN \to NN\pi\pi$ reactions for the beam momenta around 3.0
GeV~\cite{caoprc}. Up to now the properties of these resonances are
not well established and especially the nature of $N^*(1440)$ is
still in controversial~\cite{ppdata}. Therefore it is meaningful to
examine whether $\bar{N}N \to \bar{N}N\pi\pi$ reactions could supply
us with useful information. Also the $\bar{p}N \to \bar{N}N\pi\pi$
channels could serve as a complementary place to test and verify the
results of $pN\to NN\pi\pi$ reactions. As we shall demonstrate
later, some channels in antinucleon-nucleon collisions may be very
suitable to settle down the problems found in nucleon-nucleon
collisions.

Our paper is organized as the following. In Sect.~\ref{formalism},
we present the formalism and ingredients in our calculation. Then we
give our numerical results and discussion in Sect.~\ref{discussion},
and a brief summary in Sect.~\ref{summary}.

\section{FORMALISM AND INGREDIENTS} \label{formalism}

\begin{figure}[htbp]
  \begin{center}
 {\includegraphics*[scale=0.8]{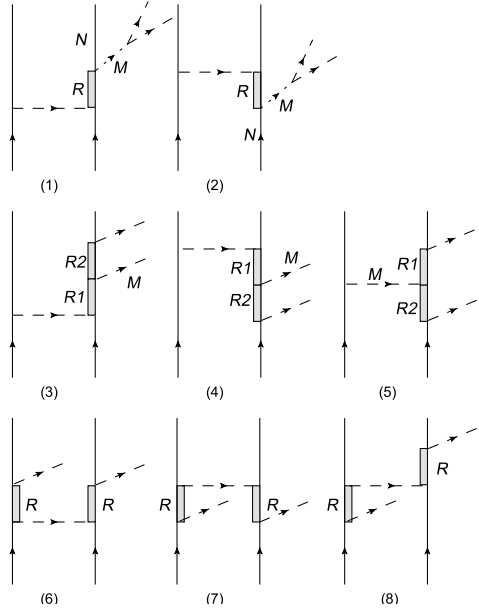}}
\caption{Feynman diagrams for $\bar{N}N\to \bar{N}N\pi\pi$. The
solid, dashed and dotted lines stand for the (anti)nucleon, mesons
and intermediate $\sigma$(or $\rho$)-meson. The shading histograms
represent the intermediate baryon resonances or off-shell
(anti)nucleon. In the text, we use $R\to NM$, $R1\to R2M$ and
double-$R$ to label (1)(2), (3)(4)(5) and (6)(7)(8), respectively.}
\label{fdg}
  \end{center}
\end{figure}

The Feynman diagrams of $\bar{N}N\to \bar{N}N\pi\pi$ we considered
are depicted in Fig.~\ref{fdg}. In the case of $NN\to NN\pi\pi$
reactions it is needed to symmetrize the initial and final nucleons
so there are additional exchanged diagrams, which do not appear in
$\bar{N}N\to \bar{N}N\pi\pi$ channels. The pre-emission diagrams are
found to be small in $NN\to NN\pi\pi$ reactions~\cite{caoprc}. Here
we include them only for completeness. The formalism and parameters
are nearly the same as those used in the study of the $NN\to
NN\pi\pi$. The commonly used Lagrangians for
Meson-(anti)Nucleon-(anti)Nucleon couplings~\cite{bonn} are,
\begin{equation}
{\cal L}_{\pi N N} = - \frac{f_{\pi N N}}{m_{\pi}} \overline{N}
\gamma_5 \gamma_{\mu} \vec\tau \cdot \partial^{\mu} \vec{\pi} N ,
\label{pin}
\end{equation}
\begin{equation}
{\cal L}_{\pi \Delta \Delta} = \frac{f_{\pi \Delta \Delta}}{m_{\pi}}
\overline{\Delta}^{\nu} \gamma_5 \gamma_{\mu} \vec{\cal K} \cdot
\partial^{\mu} \vec{\pi} \Delta_{\nu},
\end{equation}
\begin{equation}
{\cal L}_{\eta N N} = -i g_{\eta N N} \overline{N} \gamma_5 \eta N,
\label{etan}
\end{equation}
\begin{equation}
{\cal L}_{\sigma N N} = g_{\sigma N N}  \overline{N} \sigma N,
\label{sign}
\end{equation}
\begin{equation}
{\cal L}_{\rho N N} = -g_{\rho N N}
\overline{N}(\gamma_{\mu}+\frac{\kappa}{2m_N} \sigma_{\mu \nu}
\partial^{\nu})\vec\tau \cdot \vec\rho^{\mu} N. \label{rhon}
\end{equation}
In the above and following, we explicitly specify the isospin structure of the
isospin 3/2 fields. The isospin transition operators $\cal{I}$ and $\cal{K}$ are defined as,
\begin{equation}
{\cal I}_{mn} = \sum\limits_{l=0,\pm 1}(1l\frac{1}{2} n | \frac{3}{2} m)\hat{e}^*_l
\end{equation}
\begin{equation}
{\cal K}_{mn} = \sum\limits_{l=0,\pm 1}(1l\frac{3}{2} n | \frac{3}{2} m)\hat{e}^*_l
\end{equation}
where $m$ and $n$ are the third components of the isospin
projections, and $\vec\tau$ the Pauli matrices.
At each vertex a relevant off-shell form factor should be used and
we take them as~\cite{bonn},
\begin{equation}
F^{NN}_M(k^2_M)=\left(\frac{\Lambda^2_M-m_M^2}{\Lambda^2_M-
k_M^2}\right)^n,
\end{equation}
with n=1 for $\pi$- and $\eta$-meson and n=2 for $\rho$-meson.
$k_M$, $m_M$ and $\Lambda_M$ are the 4-momentum, mass and cut-off
parameters for the exchanged meson, respectively. The coupling
constants  and the cutoff parameters are taken as
~\cite{bonn,tsushima}: $f_{\pi NN}^2/4\pi = 0.078$, $g_{\eta
NN}^2/4\pi = 0.4$, $g_{\sigma NN}^2/4\pi = 5.69$,
$g_{\rho NN}^2/4\pi = 0.9$, $\Lambda_{\pi}$ = $\Lambda_{\eta}$ = 1.0 GeV,
$\Lambda_{\sigma}$ = 1.3 GeV, $\Lambda_{\rho}$ =
1.6 GeV, and $\kappa$ = 6.1. We use $f_{\pi \Delta \Delta}=4f_{\pi
NN}/5$ from the quark model~\cite{reviewlee}. The mass and
width of $\sigma$-meson are adopted as 550 MeV and 500 MeV, respectively.

In the $NN\to NN\pi\pi$ reactions we have shown that the
$N^*(1440)$, $\Delta(1232)$, $\Delta^*(1600)$ and $\Delta^*(1620)$
resonances play the major role in the considered
energies~\cite{caoprc}. Other resonances give negligible
contributions so we can safely ignore them. The effective
Lagrangians for the relevant resonance couplings
are~\cite{zoucoupling,ouyang},
\begin{equation}
{\cal L}_{\pi N\Delta} = g_{\pi N\Delta} \overline{N} \vec{\cal I} \cdot
\partial^{\mu} \vec{\pi} \Delta_{\mu} + h.c.,
\end{equation}
\begin{equation}
{\cal L}_{\pi NN^*_{(1440)}} = g_{\pi NN^*_{(1440)}} \overline{N} \gamma_5
\gamma_{\mu} \vec\tau \cdot \partial^{\mu} \vec{\pi}N^*_{(1440)} + h.c.,
\end{equation}
\begin{equation}
{\cal L}_{\sigma NN^*_{(1440)}} = g_{\sigma NN^*_{(1440)}}  \overline{N} \sigma
N^*_{(1440)} + h.c.,
\end{equation}
\begin{equation}
{\cal L}_{\pi \Delta N^*_{(1440)}} = g_{\pi \Delta N^*_{(1440)}}
\overline{\Delta}_{\mu} \vec{\cal I} \cdot \partial^{\mu} \vec{\pi} N^*_{(1440)} +
h.c.,
\end{equation}
\begin{equation}
{\cal L}_{\pi N\Delta^*_{(1620)}} = g_{\pi N\Delta^*_{(1620)}} \overline{N} \vec{\cal I} \cdot
\vec{\pi} \Delta^*_{(1620)} + h.c.,
\end{equation}
\begin{equation}
{\cal L}_{\rho N\Delta^*_{(1620)}} = g_{\rho N\Delta^*_{(1620)}}\overline{N} \gamma_5
(\gamma_{\mu}-\frac{q_{\mu} \not \! q}{q^2}) \vec{\cal I} \cdot
\vec{\rho^{\mu}} \Delta^*_{(1620)} + h.c.,
\end{equation}
\begin{equation}
{\cal L}_{\pi \Delta\Delta^*_{(1620)}} = g_{\pi \Delta\Delta^*_{(1620)}}
\overline{\Delta}_{\mu} \gamma_5 \vec{\cal K} \cdot \partial^{\mu}
\vec{\pi} \Delta^*_{(1620)} + h.c.,
\end{equation}
\begin{equation}
{\cal L}_{\pi N\Delta^*_{(1600)}} = g_{\pi N\Delta^*_{(1600)}} \overline{N} \vec{\cal I} \cdot
\partial^{\mu} \vec{\pi} \Delta^*_{(1600)\mu} + h.c.,
\end{equation}
\begin{equation}
{\cal L}_{\pi \Delta\Delta^*_{(1600)}} = g_{\pi \Delta\Delta^*_{(1600)}}
\overline{\Delta}^{\mu} \gamma_5 \vec{\cal K} \cdot \vec{\pi} \Delta^*_{(1600)\mu} +
h.c.,
\end{equation}
\begin{equation}
{\cal L}_{\pi N^*_{(1440)}\Delta^*_{(1600)}} = g_{\pi N^*\Delta^*_{(1600)}} \overline{N^*}
\vec{\cal I} \cdot
\partial^{\mu} \vec{\pi} \Delta^*_{(1600)\mu} + h.c.,
\end{equation}
For the Resonance-Nucleon-Meson vertices, form factors with the
following form are used:
\begin{equation}
F^{RN}_M(k^2_M)=\left(\frac{\Lambda^{*2}_M-m_M^2}{\Lambda^{*2}_M- k_M^2}\right)^n,\label{nmrformf}
\end{equation}
with n=1 for $N^*$ resonances and n=2 for $\Delta^*$ resonances. We employ $\Lambda^*_{\pi}$ = $\Lambda^*_{\sigma}$ = $\Lambda^*_{\eta}$ = $\Lambda^*_{\rho}$ = 1.0 for $N^*(1440)$, $\Delta(1232)$, $\Delta^*(1620)$ and $\Lambda^*_{\pi}$ = 0.8 for $\Delta^*(1600)$. The
Blatt-Weisskopf barrier factors $B(Q_{N^*\Delta\pi})$ are used in the
$N^*(1440)$-$\Delta$-$\pi$ vertices~\cite{zouform},
\begin{equation}
B(Q_{N^*\Delta\pi})=\sqrt{\frac{P^2_{N^*\Delta\pi}+Q^2_{0}}{Q^2_{N^*\Delta\pi}+Q^2_{0}}},
\end{equation}
Here $Q_0$ is the hadron scale parameter, $Q_0=0.197327/R$ GeV/c
with R the radius of the centrifugal barrier in the unit of fm and
chosen to be 1.5 fm to fit the data of $NN\to NN\pi\pi$ reactions.
$Q_{N^*\Delta\pi}$ and $P_{N^*\Delta\pi}$ is defined as,
\begin{equation}
Q^2_{N^*\Delta\pi}=\frac{(s_N^*+s_\Delta-s_\pi)^2}{4s_N^*}-s_\Delta,
\end{equation}
\begin{equation}
P^2_{N^*\Delta\pi}=\frac{(m^{2}_{N^*}+m^2_\Delta-m^2_\pi)^2}{4m^{2}_{N^*}}-m^2_\Delta,
\end{equation}
with $s_x$ being the invariant energy squared of $x$ particle. We
introduce the Blatt-Weisskopf barrier factors only for
$N^*(1440)$-$\Delta$-$\pi$ vertices because other resonances, namely
$\Delta^*(1600)$ and $\Delta^*(1620)$, begin to contribute at high
energies so these factors have little influence on their behavior at
the considered energies. On the other hand, as we have addressed,
the data of nucleon-nucleon collisions at high energies are scarce
so it is meaningful to decrease the adjustable parameters by using
fewer form factors. If we would have enough data or go to higher
energies it should be certainly necessary to include these form
factors in the model.

Because the mass of $\sigma$-meson is near the two-$\pi$ threshold,
the following Lagrangians and form factor are employed for the
$\sigma$-$\pi$-$\pi$ vertex~\cite{reviewlee,juelich},
\begin{equation}
{\cal L}_{\sigma\pi\pi} =
g_{\sigma\pi\pi}\partial^{\mu}\vec{\pi}\cdot
\partial_{\mu} \vec{\pi} \sigma,
\end{equation}
\begin{equation}
{\cal L}_{\rho\pi\pi} = g_{\rho\pi\pi}\vec{\pi}\times
\partial_{\mu} \vec{\pi}\cdot \vec{\rho^{\mu}},
\end{equation}
\begin{equation}
F^{\pi\pi}_{\sigma}(\vec{q}^{2})=\left(\frac{\Lambda^{2}+\Lambda_0^{2}}{\Lambda^{2}+
\vec{q}^{2}}\right)^2,
\end{equation}
where $\vec{q}$ is the relative momentum of the emitted pion in the
center of mass system. We use $\Lambda = 0.8$
GeV and $\Lambda_0^{2} = 0.12$
GeV$^2$ to normalize this form factor to unity when $\pi$- and
$\sigma$- meson are all on-shell. The decay width of $\sigma\to\pi\pi$
and $\rho\to\pi\pi$ yield $g^2_{\sigma\pi\pi}$ = $6.06$ and $g^2_{\rho\pi\pi}$ = $2.91$.


\begin{table}[htbp]
\caption{Relevant parameters used in our calculation. The masses,
widths and branching ratios (BR) are taken from central values of
PDG~\cite{pdg2008} except the BR for $N^*(1440)\to\Delta\pi$.}
\label{coupling}
\begin{center}
\begin{tabular}{cccccc}
\hline\hline
Resonance   &Pole Position &BW Width & Decay Mode & Decay Ratio &$g^2/4\pi$\\
\hline
$\Delta^*(1232) P_{33}$ & (1210, 100)& 118& $N\pi$& 1.0& 19.54\\
$N^*(1440) P_{11}$  & (1365, 190)& 300& $N\pi$& 0.65& 0.51\\
     &  &  & $N\sigma$& 0.075&3.20\\
     &  &  & $\Delta\pi$& 0.135&4.30\\
$\Delta^*(1600) P_{33}$  & (1600, 300)& 350& $N\pi$& 0.175& 1.09\\
     &  &  & $\Delta\pi$& 0.55&59.9\\
     &  &  & $N^*(1440)\pi$& 0.225&289.1\\
$\Delta^*(1620) S_{31}$  & (1600, 118)& 145& $N\pi$& 0.25& 0.06\\
     &  &  & $N\rho$& 0.14&0.37\\
     &  &  & $\Delta\pi$& 0.45&83.7\\
\hline\hline
\end{tabular}
\end{center}
\end{table}

The form factor for the baryon resonance R, $F_{R}(q^2)$, is taken
as,
\begin{equation}
F_{R}(q^2)=\frac{\Lambda_R^{4}}{\Lambda_R^{4} + (q^2-M^2_R)^2},\label{resonanceff}
\end{equation}
with $\Lambda_R$ = 1.0 GeV. The same type of form factors are also
applied to the nucleon pole with $\Lambda_N$ = 0.8 GeV. The
propagators of the exchanged meson, nucleon pole and resonances can
be written as~\cite{tsushima,wu},
\begin{equation}
G_{\pi/\eta}(k_{\pi/\eta})=\frac{i}{k^{2}_{\pi/\eta}-m^{2}_{\pi/\eta}},
\end{equation}
\begin{equation}
G_{\sigma}(k_{\sigma})=\frac{i}{k^{2}_{\sigma}-m^{2}_{\sigma}+im_{\sigma}\Gamma_{\sigma}},
\end{equation}
\begin{equation}
G^{\mu\nu}_{\rho}(k_{\rho})=-i\frac{g^{\mu\nu}-k_{\rho}^{\mu}
k_{\rho}^{\nu}/k_{\rho}^{2}}{k^{2}_{\rho}-m^{2}_{\rho}},
\end{equation}
\begin{equation}
G_{N}(q)=\frac{ -i(\not \! q \pm m_{N})}{q^2-m^2_{N}}.
\end{equation}
\begin{equation}
G^{1/2}_{R}(q)=\frac{ -i(\not \! q
\pm M_{R})}{q^2-M^2_{R}+iM_{R}\Gamma_{R}}.
\end{equation}
\begin{equation}
G^{3/2}_{R}(q)=\frac{ -i (\not\! q \pm M_R)
G_{\mu\nu}(q)}{q^2-M^2_{R}+iM_{R}\Gamma_{R}}.
\end{equation}
Here "$\pm$" is for particles and antiparticles, respectively.
$\Gamma_{R}$ is the total width of the corresponding resonance,
and $G_{\mu\nu}(q)$ is defined as,
\begin{equation}
G_{\mu \nu}(q) = - g_{\mu \nu} + \frac{1}{3} \gamma_\mu \gamma_\nu \pm
\frac{1}{3 M_R}( \gamma_\mu q_\nu - \gamma_\nu q_\mu) + \frac{2}{3
M^2_R} q_\mu q_\nu,
\end{equation}
Because constant width is used in the Breit-Wigner (BW) formula, we
adopt the pole positions of various resonances for parameters
appearing in the propagators.

The coupling constants appearing in relevant resonances were
determined by the empirical partial decay width of the resonances
taken from Particle Data Group (PDG)~\cite{pdg2008}, and the
detailed calculations of $g_{\rho NR}$ and $g_{\sigma NR}$ from the
$R \to N\rho(\sigma) \to N\pi\pi$ decay were given in
Ref.~\cite{xierho}. The values of cut-off in form factors were
adjusted to fit the data of $NN\to NN\pi\pi$ reactions by
hand~\cite{caoprc}. Here we would like to mention that in our fit we
first determined the cut-off values of $\Delta$ and $\Delta^*$
resonances by the $pp\to nn\pi^+\pi^+$ channel which had very small
$N^*$ contribution to be much cleaner, and then it was much easier
for us to pin down the $N^*$ contributions by a large amount of data
in $pp\to pp\pi^+\pi^-$ and $pp\to pp\pi^0\pi^0$ channels. We tried
to use the same values in the same kind of form factors for all
resonances with the aim to reduce the number of free parameters in
the model. Take the resonance form factor in Eq.~\ref{resonanceff}
for example, we employed $\Lambda_R$ = 1.0 GeV for all the
resonances. But in some of form factors, we used different cut-off
values for resonances in order to reproduce the data better. For
instance, the $\Lambda^*_{\pi}$ for $\Delta^*(1600)$ in
Eq.\ref{nmrformf} was not the same with other resonances. It should
be noted that we adopted a nearly half of the decay width of
$N^*(1440)\to \Delta\pi$ in PDG as the recent data of $NN\to
NN\pi\pi$ and $\gamma p\to p\pi^0\pi^0$ reactions
favored~\cite{ppdata,roperdecay}. The used decay width of
$N^*(1440)\to N\sigma$ is the same with the value in PDG because we
achieved an agreement with the data by adjusting the relevant
cut-off parameters. So in our model a larger decay width of
$N^*(1440)\to N\sigma$ compared to PDG was not required. The values
of coupling constants and cut-off used in our computation are
compiled in Table~\ref{coupling}, together with the properties of
the resonances and the central values of branch ratios. As we
addressed, the parameters in Table~\ref{coupling} are the same as we
used in the analysis of $NN\to NN\pi\pi$ reactions~\cite{caoprc}. So
we do not introduce any further free parameters and the calculated
results of $\bar{N}N\to \bar{N}N\pi\pi$ in fact can be viewed as the
predictions of our model.

The amplitudes can be obtained straightforwardly by applying the
Feynman rules to the diagrams in Fig.~\ref{fdg}. Isospin
coefficients are considered in different isospin channels. We do not
include the interference terms among different diagrams because
their relative phases are unknown. The Valencia model seems to show
that such terms are very small~\cite{Alvarez}, and our analysis of
$NN\to NN\pi\pi$ reactions also reproduce the data well without
including these terms. The multi-particle phase space integration
weighted by the amplitude squared can be performed by a Monte Carlo
program using the code FOWL from the CERN program
library~\cite{fowl}.

\section{Numerical RESULTS AND DISCUSSION} \label{discussion}

Fig.~\ref{tcsmajor} and Fig.~\ref{tcsminor} demonstrate our
calculated total cross sections of four isospin channels together
with the existing data~\cite{JETSET,lysdata,eastdata,lbbook}. Our
numerical results give an overall good reproduction to all three
initial $\bar pn$ channels, but overestimate the initial $\bar pp$
channel. This may be due to the fact that we have not treated the
initial state interactions (ISI) properly. The ISI usually has a
weak energy dependence for the meson production processes, so
adjusting cutoff parameters in the form factors may partly account
for it effectively as for the pp collision~\cite{caoprc,ouyang}.
However, while the $\bar pn$ is a pure isospin-vector state, the
$\bar pp$ is a mixture of isospin-scalar and isospin-vector. The
annihilation rate for the isospin-scalar $\bar NN$ is empirically
found to be bigger than the isospin-vector by a factor about
1.7~\cite{Locher}. The different annihilation rates will cause
different ISI reduction factors~\cite{ISI,Haidenbauer1992}. This
effect is not taken into account in our model calculation. The final
state interactions (FSI) may also cause smooth energy-dependent
modifications to the total cross sections~\cite{FSI}. Although the
ISI and FSI could be taken into account by some more complicated
approaches~\cite{ISI,Haidenbauer1992,FSI}, they are still of some
model dependence.  Since in this work we mainly investigate the
relative importance of various resonance contributions, we have not
included complicated treatments of ISI and FSI which are not
expected to influence our main conclusions.

In the following, we shall first address the $\bar{p}n \to
p\bar{n}\pi^-\pi^-$ channel because it is similar to the $pp \to
nn\pi^+\pi^+$ channel and has negligible $N^*$ contribution to be
more clean. Then we shall discuss other three channels. We use the
same definitions of various differential cross sections as those
used in $NN \to NN\pi\pi$ reactions~\cite{caoprc,ppdata}. The
$M_{ij}$ and $M_{ijk}$ are the invariant mass spectra, and the
angular distributions are all defined in the overall center of mass
system. The values of vertical axis in the presented figures are all
arbitrarily normalized. For concreteness we list the definitions of
the angular distributions in the following,

$\Theta_M$: the scattering angle of $M$;

$\delta_{ij}$: the opening angle between $i$ and $j$ particles;

$\Theta^{ij}_i$: the scattering angle of
$i$ in the rest frame of $i$ and $j$ with respect to the beam axis;

$\vartheta_{i}^{ij}$: the scattering angle of
$i$ in the rest frame of $i$ and $j$ with respect to the sum of momenta of $i$ and $j$,
corresponding to $\widehat{\Theta}_{i}^{ij}$
defined in Ref.~\cite{ppdata}.

We try to give adequate information to the future measurements so we
show a lot of observables predicted by our model in the following.
PANDA is expected to install a $4\pi$ solid angle detector with good
particle identification for charged particles and photons to get the
data of differential cross sections with good quality. Then if there
are any experimental results in the future we can immediately know
whether our model works and which aspect should be improved in view
of the shortcomings of our model. Taking $M_{N\bar{N}}$ and
$M_{N\pi}$ as examples, we could identify the role of final state
interaction and various resonances, respectively. The angular
distributions are also useful to identify different contributions,
especially some of which may be sensitive to the details of reaction
mechanism.

\subsection{The channel of $\bar{p}n \to p\bar{n}\pi^-\pi^-$} \label{nnpipi}

For this channel, the $N^*\to N\sigma$ and $N^*\to\Delta\pi$ cannot
contribute. The $\Delta \to N\pi\to N\pi\pi$ term is dominant for
the energies below 2000 MeV in this channel as shown in
Fig.~\ref{tcsmajor}(c) and Fig.~\ref{tcsminor}(c). Because the
$\Delta \to N\pi\to N\pi\pi$ contribution is found to be important
to describe the data of various $NN\to NN\pi\pi$ reactions
simultaneously~\cite{caoprc}, it would be very useful to find other
place to get some constrain on this term. The $\bar{p}n \to
p\bar{n}\pi^-\pi^-$ reaction is just an excellent place for such
purpose. In Fig.~\ref{nnpipi} we give the differential cross
sections at the beam momentum of 1800 MeV. The peak of invariant
mass spectrum $M_{\bar{n}\pi^-}$ is obviously different from that in
$pp \to nn\pi^+\pi^+$ channel which makes it easy for us to identify
the $\Delta \to N\pi\to N\pi\pi$ term. The steep rise of angular
distribution $\Theta_{\bar{n}}$ in forward angle is distinct from
the symmetric shape in $pp \to nn\pi^+\pi^+$. This is trivial
because amplitudes are not symmetric in the exchange of the
(anti)nucleons in antinucleon-nucleon collisions.

For the energies above 2300 MeV the $\Delta^*(1600)$ and
$\Delta^*(1620)$ terms become significant and it is a good place to
study the properties of them. The contribution from $\Delta^*(1600)
\to N^*(1440)\pi$ term begins to take over as the biggest one in
this energy region.

As pointed out in our analysis of $NN\to NN\pi\pi$, the $pp \to
nn\pi^+\pi^+$ is very crucial in determining the cut-off parameters
for the form factors of relevant $\Delta^*$ resonances due to the
fact that this channel has negligible $N^*$ contribution.
Unfortunately, the current data of differential cross sections of
$pp \to nn\pi^+\pi^+$ suffer large uncertainties~\cite{ppdata}
because among its final four particles there are two neutrons which
are difficult to detect. Especially, it is hard to figure out that
whether there is any dump hump structure or not in $M_{\pi^+\pi^+}$
from the current data of $pp \to nn\pi^+\pi^+$~\cite{caoprc}. The
channel of $\bar{p}n \to p\bar{n}\pi^-\pi^-$ is just the same as
clean as the $pp \to nn\pi^+\pi^+$ but with only one antineutron in
its final four particles, which can be easily reconstructed by the
missing mass spectrum. Another ambiguity may rise up from the
spectator proton when deuteron target is used to analyze $\bar{p}d
\to p\bar{n}p_{spec}\pi^-\pi^-$, but spectator model is repeatedly
confirmed to be reliable in (anti)nucleon-nucleon collisions. So
$\bar{p}n \to p\bar{n}\pi^-\pi^-$ reaction is strongly suggested to
be analyzed in PANDA-FAIR and it will be very helpful to distinguish
different models.

\subsection{The channel of $\bar{p}p \to \bar{p}p\pi^+\pi^-$} \label{2picharge}

\begin{figure}[htbp]
  \begin{center}
{\includegraphics*[scale=0.7]{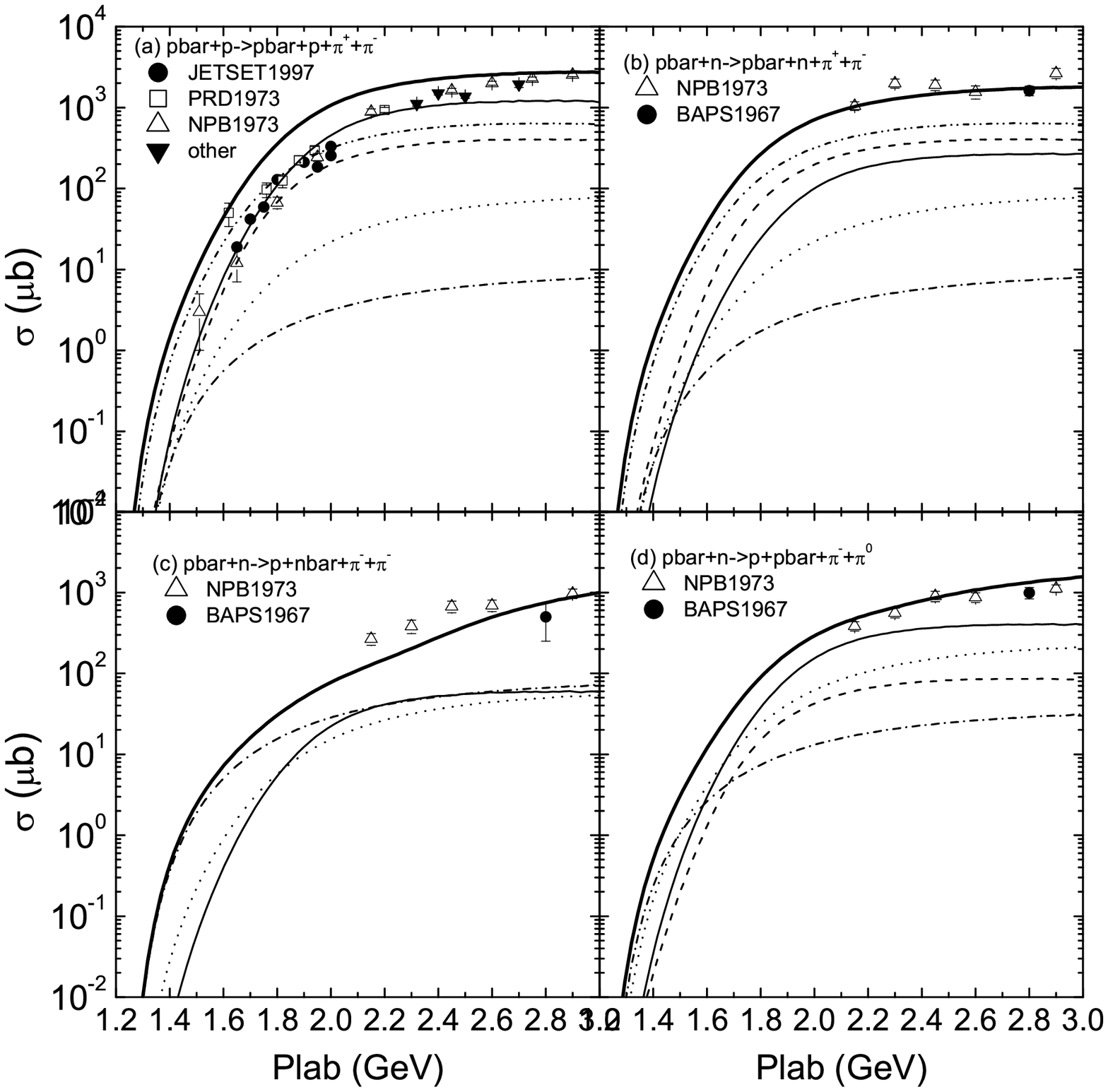}}
    \caption{Total cross sections of $\bar{N}N
    \to \bar{N}N\pi\pi$. The solid,
    dash--dot-dotted, dashed, dotted, dash-dotted, and bold
    solid curves correspond to contribution from double-$\Delta$,
    $N^*(1440) \to N\sigma$, $N^*(1440) \to \Delta\pi$, $\Delta \to \Delta\pi$,
    $\Delta \to N\pi$, and the full contributions, respectively. The data are from Refs.\protect\cite{Mason,lysdata,eastdata,lbbook,JETSET}.} \label{tcsmajor}
  \end{center}
\end{figure}

\begin{figure}[htbp]
  \begin{center}
{\includegraphics*[scale=0.7]{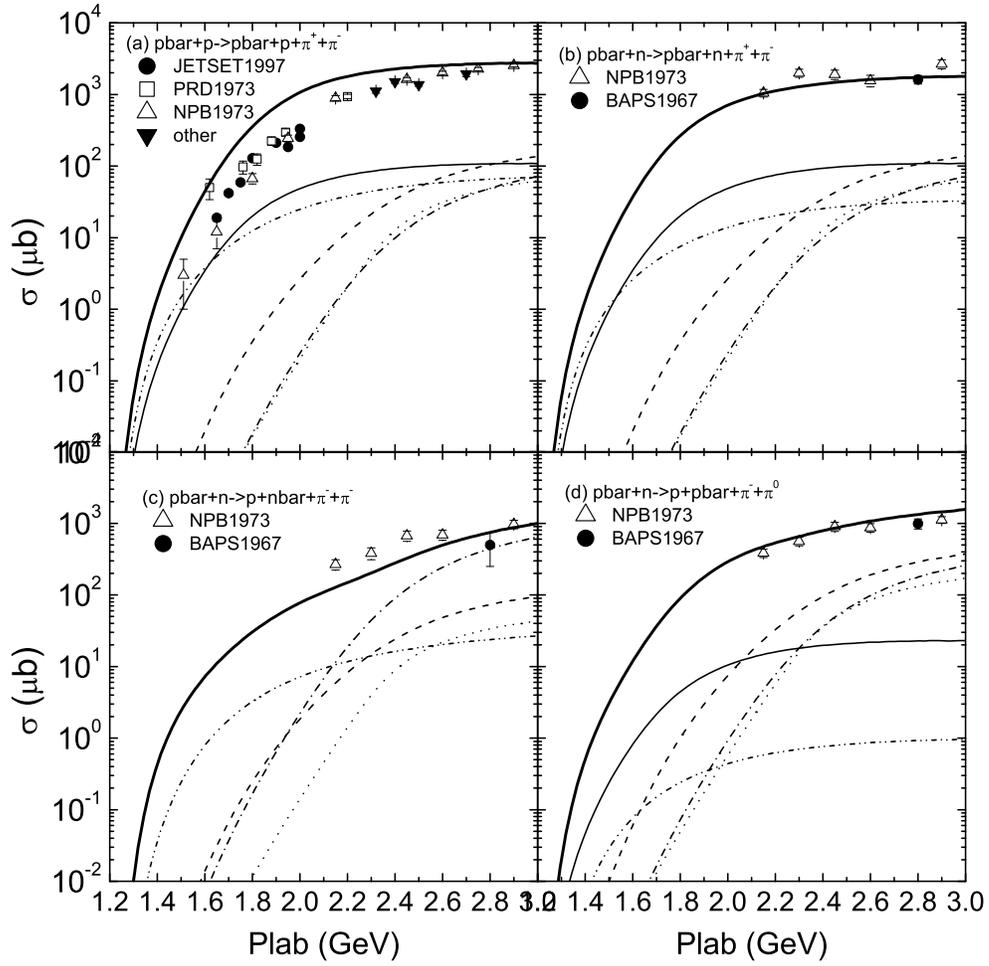}}
    \caption{Total cross sections of $\bar{N}N
    \to \bar{N}N\pi\pi$. The dashed, dash-dotted, dotted, dashed-dot-dotted,
    solid, and bold solid curves correspond to contribution from
    $\Delta^*(1600) \to \Delta\pi$, $\Delta^*(1600) \to N^*(1440)\pi$,
    $\Delta^*(1620) \to \Delta\pi$, nucleon pole, $N \to \Delta\pi$,
    and the full contributions, respectively. The data are from Refs.\protect\cite{Mason,lysdata,eastdata,lbbook,JETSET}.} \label{tcsminor}
  \end{center}
\end{figure}

This channel is interesting because its double-$\Delta$ contribution
mainly comes from the simultaneous $\bar{\Delta}^{--}$ and
$\Delta^{++}$ excitation. As depicted in Fig.~\ref{tcsmajor}(a) and
Fig.~\ref{tcsminor}(a), for the energies below 1800 MeV the
$N^*(1440) \to N\sigma$ term gives the largest contribution while
the nucleon pole and $N\to \Delta\pi$ terms also influence the
near-threshold region significantly. For the energies above 1800 MeV
the double-$\Delta$ term takes over to be the most important one
while $N^*(1440) \to N\sigma$ and $N^*(1440) \to \Delta\pi$ rank the
second and third, respectively. So unlike the $pp \to pp\pi^+\pi^-$
and $pp \to pp\pi^0\pi^0$ channels, in the whole energy region
$\bar{p}p \to \bar{p}p\pi^+\pi^-$ is not suitable to extract the
decay widths of $N^*(1440)$ because of the large double-$\Delta$
contribution. However, as the best measured channel in
antinucleon-nucleon collisions, it is useful to test models. As
shown in Fig.~\ref{tcsmajor}(a) and Fig.~\ref{tcsminor}(a), our
results overestimate the data of the $\bar{p}p \to
\bar{p}p\pi^+\pi^-$ channel for the beam momenta below 2.4 GeV. As
discussed at the beginning of this section, this may be caused by
the ISI and FSI which we have not included in our model calculation.
The effects from ISI and FSI will not influence much our estimation
of relative contributions from various intermediate baryon
resonances. This can be checked by the full phase space measurement
at PANDA/FAIR.

\begin{figure}[htbp]
  \begin{center}
{\includegraphics*[scale=1.0]{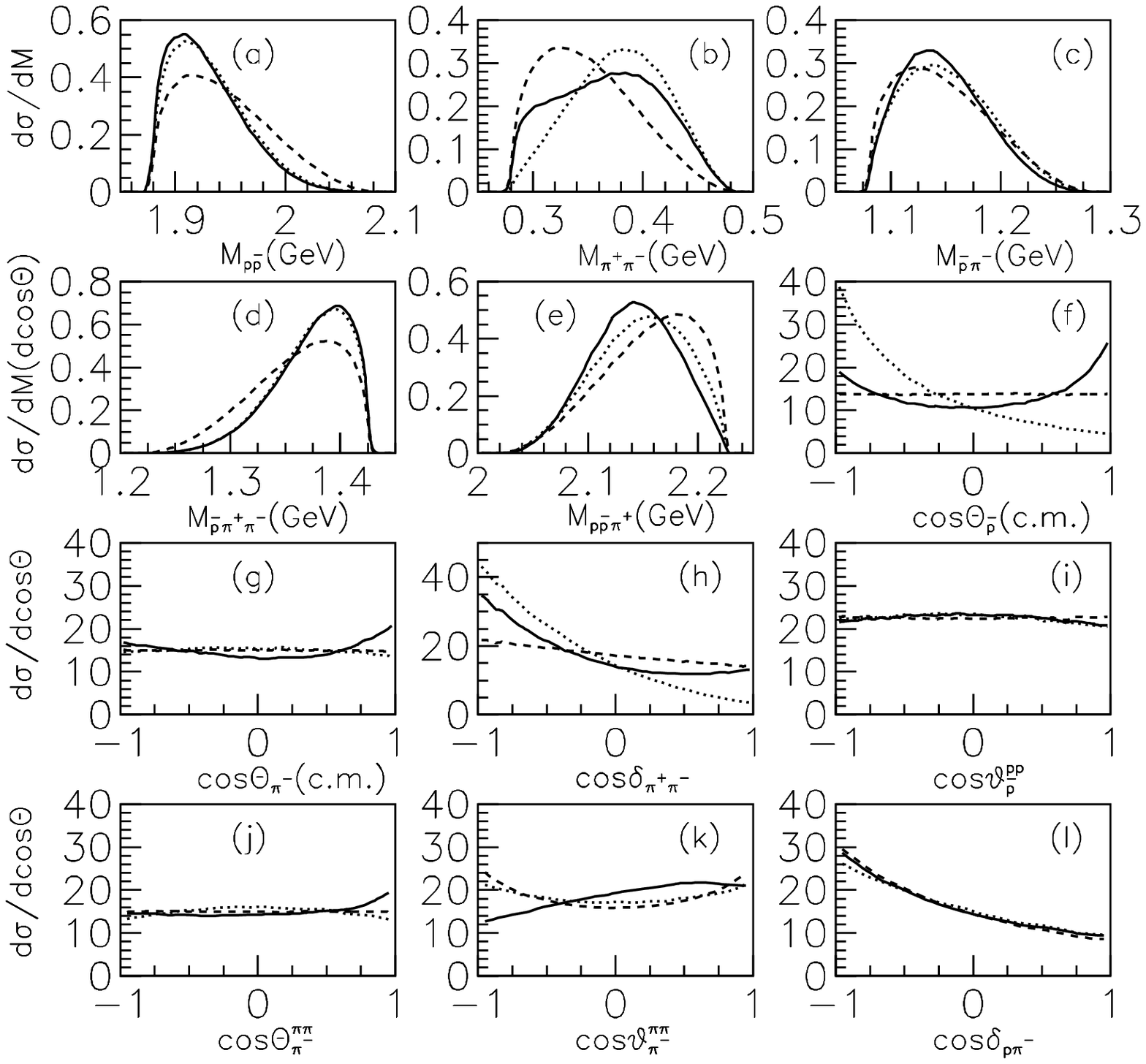}}
    \caption{Differential cross sections of $\bar{p}p \to \bar{p}p\pi^+\pi^-$
    at beam momentum 1800 MeV.  The dashed, dotted and solid curves
    correspond to the phase space, $N^*(1440)\to N\sigma$ and full
    model distributions, respectively.} \label{2picharge1100}
  \end{center}
\end{figure}

\begin{figure}[htbp]
  \begin{center}
{\includegraphics*[scale=1.0]{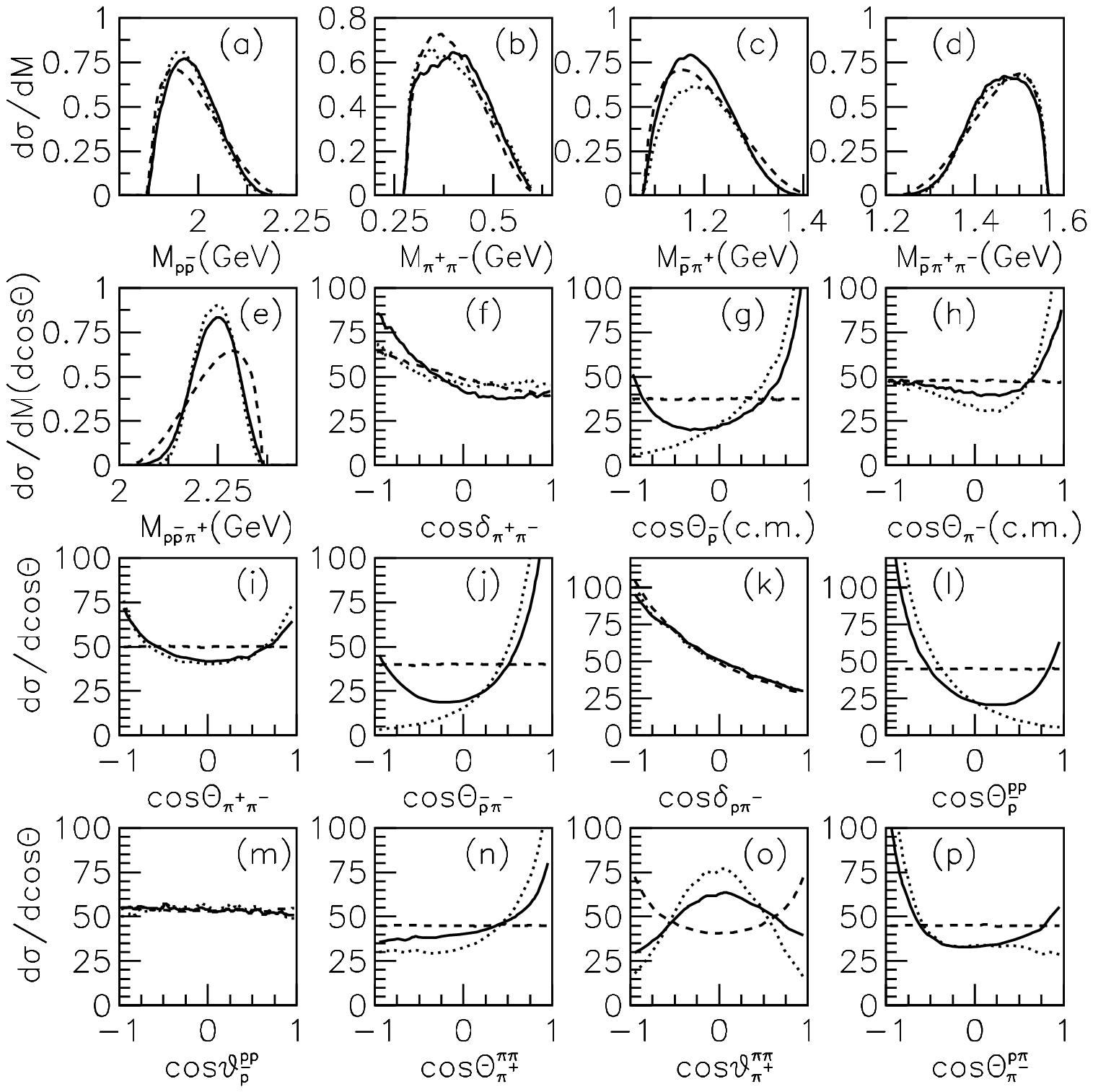}}
    \caption{Differential cross sections of $\bar{p}p \to \bar{p}p\pi^+\pi^-$
    at beam momentum 2200 MeV. The dashed, dotted and solid curves
    correspond to the phase space, double- $\Delta$ and full model
    distributions, respectively.} \label{2picharge1450}
  \end{center}
\end{figure}

In Fig.~\ref{2picharge1100} and Fig.~\ref{2picharge1450} we show the
calculated differential cross sections at the beam momenta of 1800
MeV and 2200 MeV, respectively. The $N^*(1440) \to N\sigma$,
$N^*(1440) \to \Delta\pi$ and double-$\Delta$ contributions are
comparable and have important contributions. Our results are
compatible with the old bubble chamber data measured at these
energies~\cite{lysdata}. Very similar to the $NN$ collisions, the
$\pi\pi$ system is sensitive to the change of the contributions as
can be seen in the $M_{\pi^+\pi^-}$ and
$cos\vartheta_{\pi}^{\pi\pi}$ spectrums. The double hump structure
in $M_{\pi^+\pi^-}$ caused by the $N^*(1440) \to \Delta\pi$ is
obvious. The data of $NN$ collisions did not support these
structures~\cite{caoprc} but the old bubble chamber data of
$\bar{p}p \to \bar{p}p\pi^+\pi^-$ gave obvious double hump in
$M_{\pi^+\pi^-}$ spectrums, especially at the beam momentum of about
1800 MeV~\cite{lysdata}. Unfortunately, the statistics was very low
and the number of selected events for each beam momentum was at most
several hundreds, so the measured results were inconclusive. On
theoretical side, the interference terms between $N^*(1440)\to
N\sigma$ and $N^*(1440)\to \Delta\pi$ might be relevant because
their role on the $\pi\pi$ invariant mass distributions have been
found in $\pi N\to \pi\pi N$~\cite{kamanoprc} and $NN\to
d\pi\pi$~\cite{alvarezplb}, so these terms should be treated with
care in the future work. It should be mentioned that this problem
may be related to the ABC effect of double-pion production in
nuclear fusion reactions~\cite{alvarezplb}, so it is meaningful to
extensively study it both experimentally and theoretically. The
luminosity of PANDA/FAIR is high enough to get the required
production rates so the unsettled problem of the $\pi\pi$ system can
be further explored in $\bar{p}p \to \bar{p}p\pi^+\pi^-$ channel.

\subsection{The channel of $\bar{p}n \to \bar{p}n\pi^+\pi^-$} \label{2pineutron}

In Fig.~\ref{tcsmajor}(b) and Fig.~\ref{tcsminor}(b), the $N^*(1440)
\to N\sigma$ term is found to dominate in the whole considered
energies and the nucleon pole term also gives significant
contribution in the near-threshold region. It is worth to point out
that unlike the $pn \to pn\pi^+\pi^-$ channel, the isovector
excitation of $N^*(1440)$ in $\bar{p}n \to \bar{p}n\pi^+\pi^-$ is
not enhanced compared to the $\bar{p}p\to \bar{p}p\pi^+\pi^-$
because charged meson exchange is not allowed in both channels. So
in a wide energy region it is suitable to explore the isoscalar
excitation of $N^*(1440)$.

\begin{figure}[htbp]
  \begin{center}
{\includegraphics*[scale=1.0]{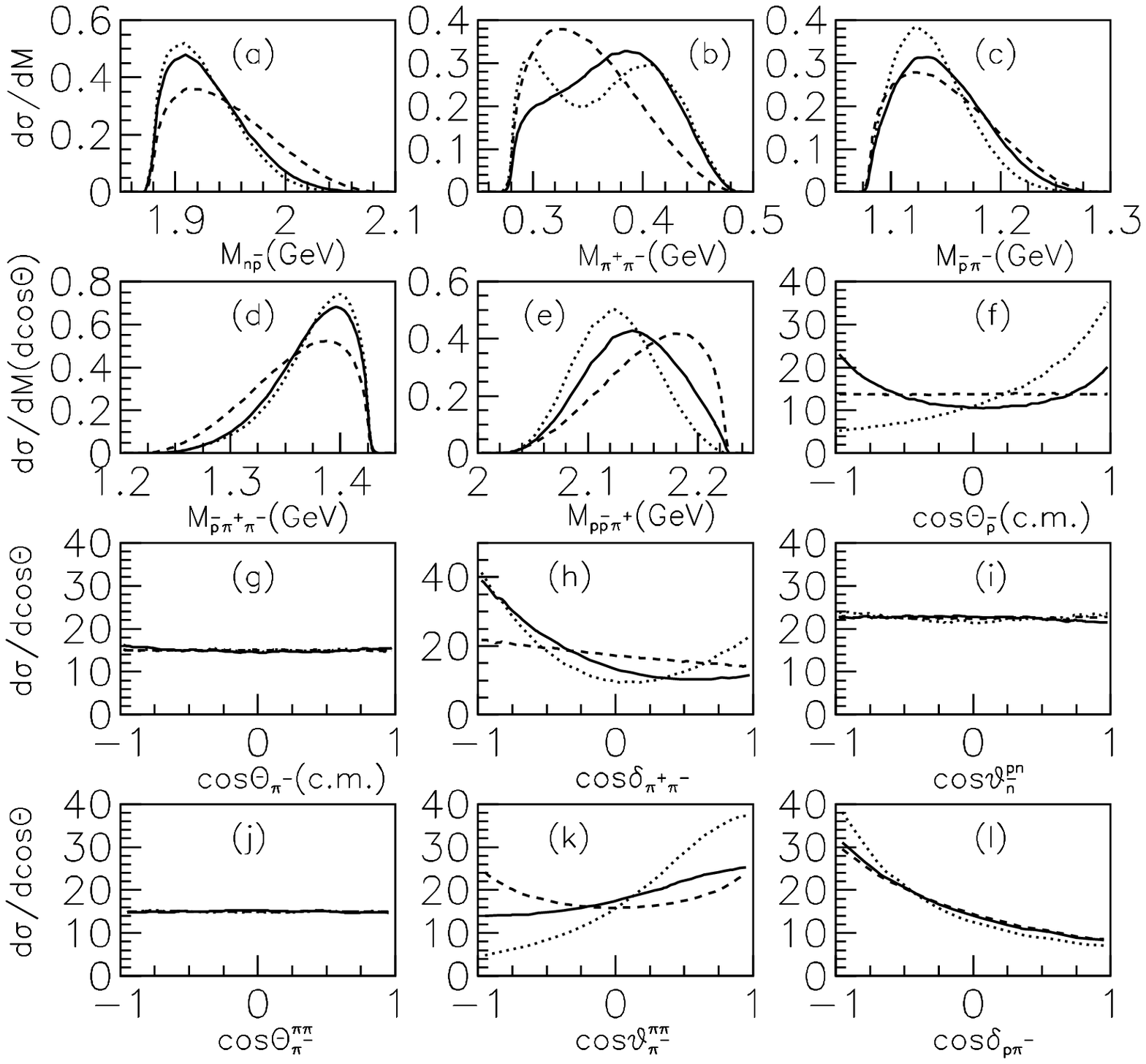}}
    \caption{Differential cross sections of $\bar{p}n \to \bar{p}n\pi^+\pi^-$
    at beam momentum 1800 MeV. The dashed, dotted and solid curves
    correspond to the phase space, $N^*(1440)\to \Delta\pi$ and full
    model distributions, respectively.} \label{pn2pi1100}
  \end{center}
\end{figure}

The $N^*(1440) \to \Delta\pi$ term is the second largest for the
beam momenta above 1500 MeV and other contributions are much
smaller. As can be seen in Fig.~\ref{pn2pi1100}, the angular
distributions of $\vartheta_{\pi^-}^{\pi\pi}$ and $\Theta_{\bar{p}}$
at 1800MeV are sensitive to the presence of $N^*(1440) \to
\Delta\pi$ term. Though there is possible ambiguity from the
spectator proton when deuteron target is used in the experiment,
this channel can be a better place to determine the partial decay
widths of $N^*(1440)$ than $pp \to pp\pi^+\pi^-$ and $pp \to
pp\pi^0\pi^0$ where they are complicated by other contributions, such as the
double-$\Delta$ and nucleon pole terms~\cite{caoprc}.

\subsection{The channels of $\bar{p}n \to \bar{p}p\pi^-\pi^0$\label{pntopp}}

The $N^*(1440) \to N\sigma$ term is not present in this reaction
because the $\sigma$-meson cannot decay to $\pi^-\pi^0$. The
double-$\Delta$ term is the most important one in a wide energy
range as shown in Fig.~\ref{tcsmajor}(d) and
Fig.~\ref{tcsminor}(d). The $\Delta \to \Delta\pi$ and
$\Delta \to N\pi\to N\pi\pi$ terms have significant contribution
below 1600 MeV and also have some contribution at higher energies
together with the $\Delta^*(1600)$ and $\Delta^*(1620)$ terms. The
agreement with the data is very good.

\subsection{The channels of $\bar{p}p \to \bar{p}p\pi^0\pi^0$,
$\bar{p}n \to \bar{p}n\pi^0\pi^0$, $\bar{p}p \to
\bar{n}p\pi^-\pi^0$\label{ppto2pi0}}

There are no data on these three channels yet. They can be measured
by PANDA experiment. The amplitudes for the $\bar{p}p \to
\bar{p}p\pi^0\pi^0$ and $\bar{p}n \to \bar{p}n\pi^0\pi^0$ channels
are the same except for the difference of $\bar pp$ and $\bar pn$
FSI. A simultaneous measurement of these
two channels may help us to understand the $\bar pp$ and $\bar pn$
FSI. These two reactions are also similar to the $pp\to
pp\pi^0\pi^0$ reaction except for a different $pp$ FSI and
interference between amplitudes by various meson exchanges.
Similarly, the $\bar pp\to\bar np\pi^-\pi^0$ reaction has an analogous
amplitude with $pn\to pp\pi^-\pi^0$ reaction except for the $\bar np$
and $pp$ FSI and interference between amplitudes.
A simultaneous study of these $\bar pN$ and $pp$
reactions may help to pin down contributions from various meson
exchanges.

\section{Summary} \label{summary}

In this paper, we present an analysis of four isospin channels of
the $\bar p N\to \bar NN\pi\pi$ reactions for the beam momenta of up
to 3.0 GeV within an effective Lagrangian approach. The model
parameters determined from the $NN \to NN\pi\pi$ reactions are used
without introducing any further free parameters. We include
contributions from the $N^*(1440)$, $\Delta$, $\Delta^*(1600)$,
$\Delta^*(1620)$ and nucleon pole to give a reasonably explanation
of the measured total cross sections. The role of the $N^*(1440)$,
$\Delta^*(1600)$ and $\Delta^*(1620)$ resonances in $\bar{N}N\to
\bar{N}N\pi\pi$ reactions have never been explored in previous
studies. We give some typical differential cross sections which can
be tested in the future measurements. We stress that PANDA
(anti-Proton ANnihilation at DArmstadt) Collaboration at the GSI-FAIR
(Facility of Antiproton and Ion Research) could play an
important role in the baryon spectrum and a large amount of events
on final states with baryon and antibaryon should be analyzed to
extract the properties of relevant resonances. The conclusions
reached from our model would be helpful to the future experiments
performed at PANDA/FAIR.

\begin{acknowledgments}

We thank J. Haidenbauer and C. Hanhart for helpful discussions. This
work was supported by the National Natural Science Foundation of
China (Nos. 10635080, 10875133, 10821063, 10925526).

\end{acknowledgments}

\begin{figure}[htbp]
  \begin{center}
{\includegraphics*[scale=1.0]{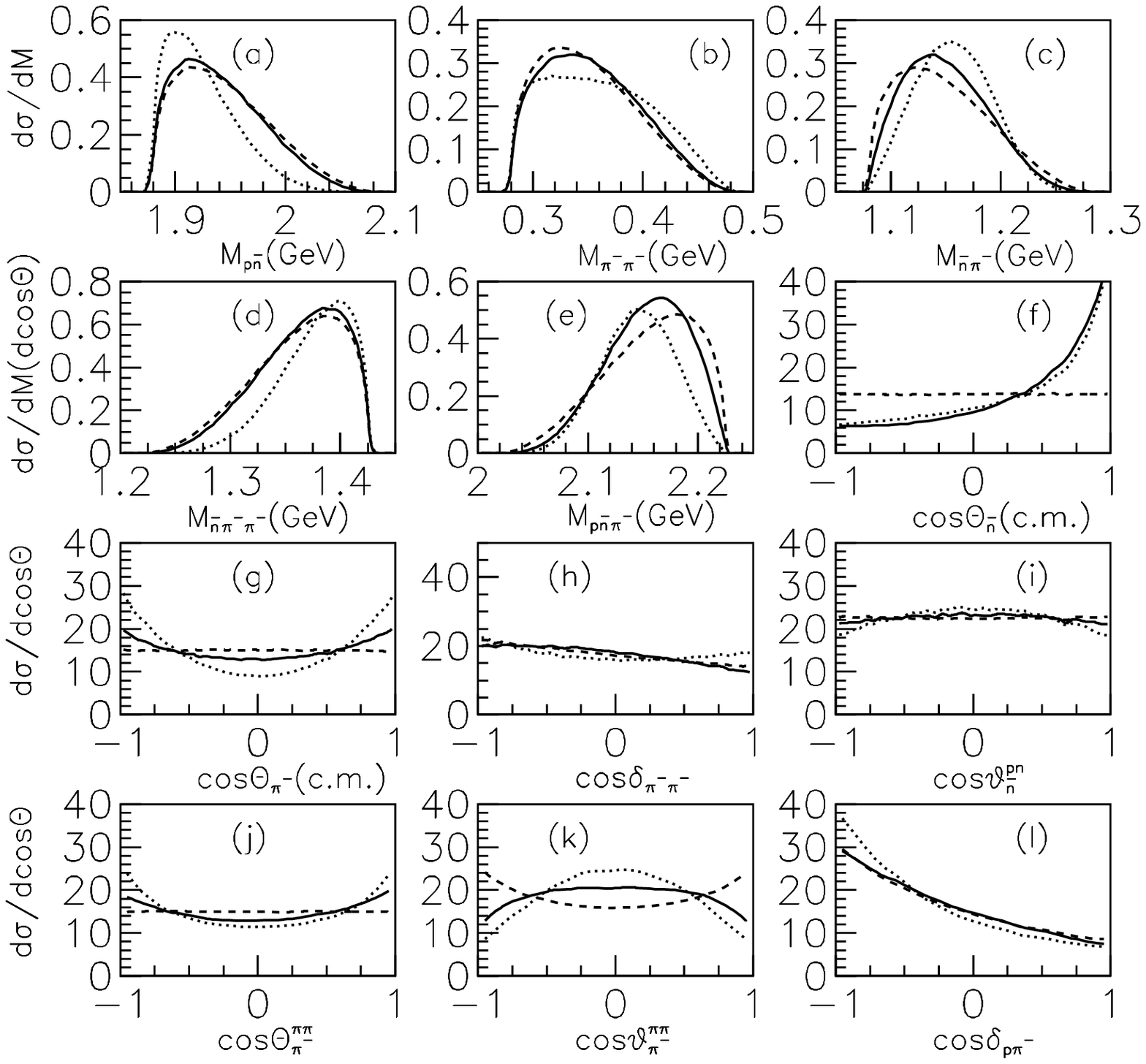}}
    \caption{Differential cross sections of $\bar{p}n \to p\bar{n}\pi^-\pi^-$
    at beam momentum 1800 MeV. The dashed, dotted and solid curves
    correspond to the phase space, double-$\Delta$ and full model
    distributions, respectively.} \label{nnpipi}
  \end{center}
\end{figure}
\end{document}